\documentclass[aps, prl, twocolumn, showpacs, superscriptaddress, floatfix, citeautoscript, superscriptaddress]{revtex4}

\usepackage[latin1]{inputenc}

\usepackage[english]{babel}

\usepackage{ragged2e}

\usepackage[final]{graphicx}

\usepackage[centertags, sumlimits, intlimits, namelimits, fleqn]{amsmath}

\usepackage{amssymb}

\usepackage{dcolumn}

\usepackage{natbib}

\usepackage{bm}

\newcommand{\qav}[1]{\langle {#1} \rangle}

\begin{document}

\title{Counting statistics for electron capture in a dynamic quantum dot}

\date{\today}

\author{Lukas \surname{Fricke}}\affiliation{Physikalisch-Technische
Bundesanstalt, Bundesallee 100, 38116 Braunschweig, Germany.}\noaffiliation

\author{Michael \surname{Wulf}}\affiliation{Physikalisch-Technische
Bundesanstalt, Bundesallee 100, 38116 Braunschweig, Germany.}\noaffiliation

\author{Bernd \surname{Kaestner}}\email[Mail to:
]{bernd.kaestner@ptb.de}\affiliation{Physikalisch-Technische
Bundesanstalt, Bundesallee 100, 38116 Braunschweig, Germany.}\noaffiliation

\author{Vyacheslavs Kashcheyevs}\affiliation{Faculty of Computing,
University of Latvia, Riga LV-1586, Latvia}\affiliation{Faculty of Physics and Mathematics,
University of Latvia, Riga LV-1002, Latvia}\noaffiliation

\author{Janis Timoshenko}\affiliation{Faculty of Computing,
University of Latvia, Riga LV-1586, Latvia}\affiliation{Faculty of Physics and Mathematics,
University of Latvia, Riga LV-1002, Latvia}\noaffiliation

\author{Pavel Nazarov}\affiliation{Faculty of Computing,
University of Latvia, Riga LV-1586, Latvia}\affiliation{Faculty of Physics and Mathematics,
University of Latvia, Riga LV-1002, Latvia}\noaffiliation

\author{Frank \surname{Hohls}}\affiliation{Physikalisch-Technische
Bundesanstalt, Bundesallee 100, 38116 Braunschweig, Germany.}\noaffiliation

\author{Philipp \surname{Mirovsky}}\affiliation{Physikalisch-Technische
Bundesanstalt, Bundesallee 100, 38116 Braunschweig, Germany.}\noaffiliation

\author{Brigitte \surname{Mackrodt}}\affiliation{Physikalisch-Technische
Bundesanstalt, Bundesallee 100, 38116 Braunschweig, Germany.}\noaffiliation

\author{Ralf \surname{Dolata}}\affiliation{Physikalisch-Technische
Bundesanstalt, Bundesallee 100, 38116 Braunschweig, Germany.}\noaffiliation

\author{Thomas \surname{Weimann}}\affiliation{Physikalisch-Technische
Bundesanstalt, Bundesallee 100, 38116 Braunschweig, Germany.}\noaffiliation

\author{Klaus \surname{Pierz}}\affiliation{Physikalisch-Technische
Bundesanstalt, Bundesallee 100, 38116 Braunschweig, Germany.}\noaffiliation

\author{Hans W. \surname{Schumacher}}\affiliation{Physikalisch-Technische
Bundesanstalt, Bundesallee 100, 38116 Braunschweig, Germany.}\noaffiliation




\begin{abstract}

We report non-invasive single-charge detection of the full probability distribution $P_n$ of the initialization of a quantum dot with $n$ electrons for rapid decoupling from an electron reservoir. We analyze the data in the context of a model for sequential tunneling pinch-off, which has generic solutions corresponding to two opposing mechanisms. One limit considers sequential ``freeze out'' of an adiabatically evolving grand canonical distribution, the other one is an athermal limit equivalent to the solution of a generalized decay cascade model. We identify the athermal capturing mechanism in our sample, testifying to the high precision of our combined theoretical and experimental methods. The distinction between the capturing mechanisms allows to derive efficient experimental strategies for improving the initialization.

\end{abstract}

\maketitle

The fast formation of quantum dots (QDs) out of a two-dimensional electron system (2DES) constitutes an open problem within the field of nanoscale electronics~\cite{Averin1991}.
The initialization process in these \emph{dynamic} QDs is a key ingredient in, e.g., devices for quantum information processing~\cite{Barnes2000, gumbs2009}, single-electron current sources~\cite{Shilton1996, blumenthal2007a, Pekola}, or nanoelectronic circuits~\cite{amakawa2004a, ono2005}.
The outcome of the initialization is characterized by a probability distribution $P_n$ for  trapping exactly $n$ electrons in the QD.
The goal is to attain a
predictable low dispersion
distribution
thus making dynamic QDs reliable and reproducible sources of electrons on demand. Deviations from this ideal case may be caused, for instance, by backtunneling~\cite{fujiwara2008, kaestner2009a, kaestner2010a} or non-adiabatic excitations~\cite{liu1993, Flensberg1999, kataoka2011}.

A decay cascade model~\cite{kaestner2010a} has been proposed recently to predict $P_n$ in dynamic QDs. It has become popular for benchmarking
QD-based current sources~\cite{giblin2010a, hohls2012, giblin2012, Fletcher2012} by extracting $P_n$ from the average current as function of QD energy using the model.
Alternative mechanisms, such as sudden decoupling from thermal equilibrium~\cite{yamahata2011} have been proposed.
Experimental distinction between the capturing mechanisms is the key towards systematic improvement of the initialization precision.
So far $P_n$ has not been measured with sufficient accuracy to allow this distinction.
The first two cumulants of $P_n$ have been extracted from current and noise measurements~\cite{Robinson2005, maire2008, Bocquillon2012}. Furthermore,
single charge detection~\cite{kataoka2007, yamahata2011} has been used to determine partial information on the distribution of QD population-depopulation events.

Here we present non-invasive charge detection to measure 
the full probability distribution $P_n$ of capturing $n$ electrons in a dynamic QD.
Considering the integer charge on the QD to be
the only degree of freedom out of equilibrium, we
derive theoretically two generic 
limits for $P_n$: a (generalized) frozen grand canonical distribution and a rate-driven athermal limit (generalizing the decay cascade model \cite{kaestner2010a}).
In the generalized form both limits may be hard to distinguish experimentally for $P_n \approx 1$, which is the relevant regime for most applications.
Yet, our experimental data for $P_n$ allow to distinguish both limits and to
conclude that the 
dynamic QD initialization 
is consistent with the 
athermal distribution.
Based on these findings, strategies for optimum high fidelity initialization are presented.

\begin{figure}[t!] \centering
\centering \includegraphics{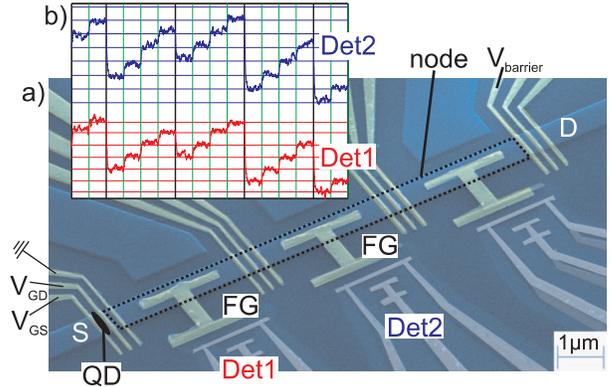}
\caption{
(Color online) (a) False-color electron microscopy image of a device.
The upper half shows the semiconducting part consisting of an 800$\,$nm wide channel (light blue) crossed by topgates (yellow). The QD is formed by the leftmost group of topgates, between source (S) and drain (D). The light-gray parts in the lower part form the SETs labeled Det1 (red) and Det2 (blue), respectively. (b) Detector signals of the charge transfer sequence, as explained in the main text.
}\label{fig:device}\end{figure}

The device under investigation is
shown in Fig.~\ref{fig:device}(a). The top half of the false-colored image shows four QD structures in series consisting each of three gates crossing a 2DES within a AlGaAs/GaAs heterostructure.
The 2DES is located 90 nm below the surface, the wet-etched channel is 800 nm wide. Similar QD structures have previously been used as single-electron current sources~\cite{blumenthal2007a, Kaestner2007c}. We use the left QD as the dynamic QD, which captures electrons from source (S) and afterwards emits them to the \emph{node} (dotted region) for charge detection. The voltage on one gate of the rightmost QD, $V_\mathrm{barrier}$, controls the transparency of the node to the drain lead (D). All other gates are grounded and do not affect the circuit. 
Close to the node two single-electron transistors (SET) based on Al--AlO$_x$--Al tunnel junctions are placed as charge detectors (Det1/Det2).
To increase the coupling between the potential of the node and the metallic detectors both are capacitively coupled by H-shaped floating gates (FG).
Correlating the detector signals allows to distinguish the electron signal from background charge fluctuations, as both detectors are coupled to the same island. All measurements are performed in a dry dilution cryostat at nominal temperature of about 25 mK.

\begin{figure}[t!] \centering
\centering \includegraphics{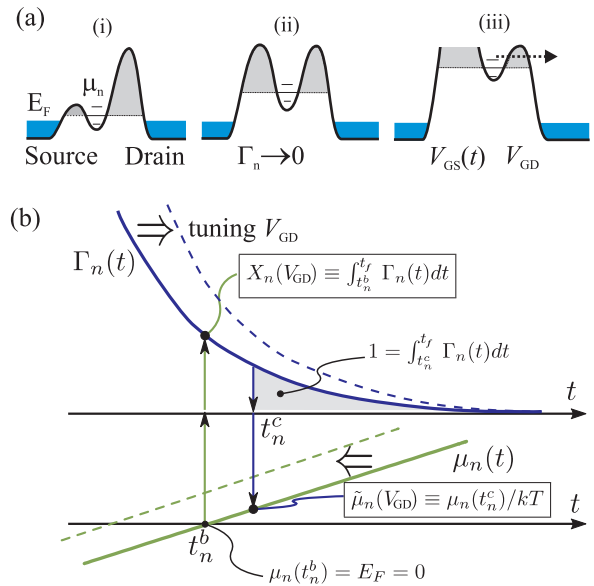}
\caption{
(Color online)  (a) Sequence of schematic potential landscapes during capturing from source (i), isolation (ii) and emission to drain (iii), respectively. Shading indicates tunneling rates $\Gamma_n$, related to area under curve. (b) Evolution of the tunneling rate $\Gamma_{n}$ and energy $\mu_{n}$ of a particular charge state for two different $V_\mathrm{GD}$ voltage settings.
}\label{fig:principle}\end{figure}

Fig.~\ref{fig:principle}(a) shows the sequence of QD initialization and charge transfer to the node schematically. An isolated QD (ii) is formed between the two left-most gates by applying sufficiently negative dc voltages, $V_{GS}$ and $V_\mathrm{GD}$ (see Fig.~\ref{fig:device}). Initialization (i) of the QD is achieved by applying the first half cycle of a sinusoidal pulse superimposed onto the source gate, so that the source barrier becomes transparent. During the subsequent rise of the source barrier a certain charge state of the isolated QD (ii) with $n$ electrons is established with probability $P_n$. In the second half cycle of the sinusoidal pulse (iii) the left barrier is raised further. As the source barrier also couples to the QD potential, i.e. acts as plunger gate~\cite{Kaestner2007c}, one can ensure complete unloading of the QD charge to the node where it can be detected non-invasively.

The cycle is repeated three times during which charges accumulate on the node. Afterwards, opening the right barrier resets the node charge, before the next three cycles start.
Example traces of the SET signals are shown in Fig.~\ref{fig:device}(b).
The bold black vertical lines represent resets of the node's charge state. The stochastic nature of this process is represented by the different initial states after each reset. Each thin green vertical line represents a combined charge capture and transfer pulse. Due to the discreteness of node charges, we can assign levels (horizontal lines) derived from a histogram of the detector trace to each interval between pulses. These are then compared to extract the number of captured electrons in this specific cycle (see Appendix).
Due to the limited bandwidth of the SET detectors ($f_\text{bw}\approx 600$ Hz) the pulses are delayed by 40 ms each. The pulse itself consists of a single period of a sine with frequency $f_\text{pump}\approx 40$ MHz.

\begin{figure}[tbh] \centering
\centering \includegraphics{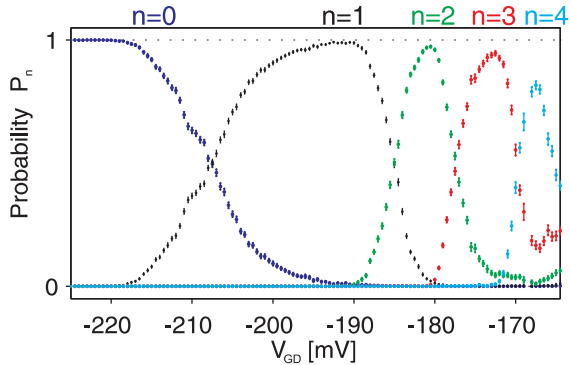}
\caption{(color online) Probabilities to capture $n=0...4$ electrons per cycle derived by counting as a function of $V_\mathrm{GD}$. The error bars indicate the 95~\% confidence interval.
}\label{fig:probs}\end{figure}

The voltage $V_\mathrm{GD}$ allows to adjust the depth of the QD potential and thereby to tune the average number $\qav{n}=\sum_{n} n P_n$ of captured electrons~\cite{kaestner2008}. Fig.~\ref{fig:probs} shows $P_n$ as measured by charge detection as function of $V_\mathrm{GD}$. The probabilities of charging the QD with up to 4 electrons are well resolved. For $n=1$ the initialization accuracy reaches 99.1~\% for $V_\mathrm{GD} \approx -192.5\,$mV. The probability to charge the QD with 4 electrons with one initialization pulse approaches 80~\% for $V_\mathrm{GD} \approx -168\,$mV.
We will analyse the $V_\mathrm{GD}$ dependence of $P_n$ further below. In the following the theoretical framework is established allowing later to relate the measured distribution to the underlying capturing mechanisms.

Treating tunneling perturbatively in a Markov approximation~\cite{AverinKorotkov1991,beenakker1PBI}, the  ``disequilibration'' and eventual freezing of $P_n(t)$ can be described by a general master equation:
\begin{align} \label{eq:det:kin1}
  \dot{P}_n(t)\!=\!P_{n\!-\!1}^{}(t)W_{n-1}^{+}(t)\!-\!P_n(t) W_{n}^{-}(t)\!+ \nonumber \\
  \!P_{n\!+\!1}^{}(t)W_{n+1}^{-}(t)\!-\!P_n^{}(t)W_{n}^{+}(t) \, ,
\end{align}
where $W^{\pm}_{n}$ are the instantaneous rates for adding ($+$) or removing ($-$) an electron to/from the QD,
averaged  over all degrees of freedom except $n$. The balance between adding and removing
defines (possibly non-equilibrium) electrochemical potentials $\mu_n(t)$ of  a QD state with $n$ electrons:
\begin{align} \label{eq:detbalmacro}
  e^{\beta \mu_n(t)} \equiv W^{-}_{n}(t)/W^{+}_{n-1}(t) \, ,
\end{align}
with $\beta^{-1} \equiv kT$ being the product of temperature and Boltzmann's constant. If the time-dependence of rates $W^{\pm}_{n}(t)$ is quasistatic, then Eq.~\eqref{eq:detbalmacro} is the expression of thermodynamic detailed balance
and $\mu_n=\Omega_n-\Omega_{n-1}$ will be set by the differences of the  thermodynamic potentials \cite{beenakker1PBI} $\Omega_n \equiv \mathcal{F}_n+E_n-n E_F$, where
$\mathcal{F}_n$ is the canonical free energy of the internal degrees of freedom on the QD,
$E_n$ is the electrostatic interaction energy and  $E_F \equiv 0$ is the Fermi level in the source lead.
Defining the total rate for charge exchange in $n\!\leftrightarrow\!n\!-\!1$ transition as
$\Gamma_n(t) \equiv W^{-}_{n}(t)+W^{+}_{n-1}(t)$ allows us to write Eq.~\eqref{eq:det:kin1}
in the form
\begin{align} \label{eq:generalKinetic}
  \dot{P}_n(t) & = -\Gamma_n \left   [\bar{f}(\mu_n) P_n(t) -f(\mu_n)  P_{n-1}(t) \right ]  + \nonumber \\
  & \Gamma_{n+1} \left [\bar{f}(\mu_{n+1})  P_{n+1}(t)- f(\mu_{n+1}) P_n(t) \right ] \, ,
\end{align}
where $f(x)=1/(1+e^{\beta x})$ is the Fermi distribution in the source and $\bar{f}\equiv 1-f$.

We will now apply Eq.~\eqref{eq:generalKinetic} to the decoupling process sketched in Fig.~\ref{fig:principle}(a-i)-(a-ii). The end of the decoupling stage (ii) is characterized by $\Gamma_{n}(t_f) = 0$. The exact time-dependencies of  $\{ \mu_n(t) \}$ and $\{ \Gamma_n(t) \}$ between an initial time moment $t_0$ and $t_f$ are impossible to predict without a more specific microscopic model for the QD.  However, two sets of critical parameters can be unambiguously defined. For each transition between $n-1$ and $n$ charges on the dot we identify two time moments: one for the onset of backtunneling, $ t_n^{b}$, and the other one for decoupling (detailed balance breakdown), $t_n^{c}$, described in Fig.~\ref{fig:principle}(b). It shows schematically the evolution of $\Gamma_n$ (blue) and $\mu_n$ (green) with time for the initialization process with two different values of $V_\mathrm{GD}$. The time for onset of backtunneling is defined by the crossing of the Fermi level, $\mu_n(t_n^{b}) =0$, while the decoupling time is set by the average number of remaining tunneling events being of order one, $\int_{t_n^{c}}^{t_f} \Gamma_n(t) d t =1$ (shaded area under the curve). Positive charging energies and the raising bottom of the confining potential well imply $t_{n+1}^{b}<t_n^{b}$ while in the relevant regime we expect the higher $n$ states to be less stable, $\Gamma_{n+1}(t) > \Gamma_{n}(t)$ (at least for $t>t_n^{b}$) and therefore generally $t_{n+1}^{c}  > t_n^{c}$. Sufficiently long equilibration before the decoupling implies that $t_n^{c}$ is well-defined for all $n$ for which the Coulomb blockade holds.
Fluctuations in the capture probability (i.e. $P_n$) will be strongest for $n$ that have $t_n^{c}$ and $t_n^{b}$ close to each other \cite{slava2012}.

We proceed by solving Eq.~\eqref{eq:generalKinetic} for $P_n \equiv P_n(t\to t_f)$  in two limits (thermal versus athermal) that correspond to opposing physical mechanisms
of charge capture. 
We assume that initially the charge on the QD is equilibrated, corresponding to a grand canonical distribution, $P_n(t_0)\propto e^{-\beta \Omega_n(t_0)}$. As long as $\Gamma_n(t)$ remain
sufficiently large, the solution $P_n(t)$ at $t>t_0$ closely follows the instantaneous equilibrium. This can be seen directly from Eq.~\eqref{eq:generalKinetic}: large $\Gamma_n$ pin
the terms in square brackets to zero and the evolving distribution of $n$ obeys detailed
balance adiabatically, 
\begin{align}\label{eq:microcanoncal}
  P_{n}(t) \approx e^{\beta \mu_{n+1}(t)} P_{n+1}(t) \, .
\end{align}
In deriving the thermal limit we consider  \emph{sudden decoupling}, i.e. $\Gamma_{n+1}(t)$ dropping to $0$ so fast that Eq.~\eqref{eq:microcanoncal} must
hold up to $t=t_{n+1}^c$ but once $t>t_{n+1}^c$ the r.h.s. of Eq.~\eqref{eq:generalKinetic}  is effectively zero
and $P_n(t)$ freezes (i.e. remains constant). With this sudden approximation, the asymptotic value $P_{n} =P_{n}(t_{n+1}^c)$ is set by a ``curtailed''
grand canonical distribution that excludes already frozen charge states with $n'<n$ but is normalized over the remaining states with $n'\ge n$
that keep being connected until $t_{n+2}^c$. 
This gives $P_n =(1-\sum_{m=0}^{n-1} P_m) \mathcal{Z}_{n+1}^{-1}$ or explicitly
\begin{align}
\label{eq:frozengrand}
   P_n & = \mathcal{Z}_{n+1}^{-1} \prod_{m=1}^{n} (1-\mathcal{Z}_m^{-1}) \, , 
\end{align}
where $\mathcal{Z}_n \equiv 1+\sum_{m=n}^{\infty} \prod_{l=n}^{m}e^{-\beta \mu_{l}(t_n^c)}=1+e^{-\beta \mu_n(t_n^c)} (1+e^{-\beta \mu_{n+1}(t_n^c)} (1+\ldots ))$
includes the electrochemical potentials $\mu_{n'}$ of states $n'\ge n$ taken at the decoupling moment $t_n^c$ of the state $n$. The assumption of well-pronounced Coulomb blockade implies that the addition energy, $\Delta \mu(t)\equiv  \mu_n(t) -\mu_{n-1}(t)$ remains large compared to temperature. Thus there exists sufficently large  $\delta_T$ such that $\beta \Delta \mu(t)  \ge \delta_T$ for all relevant $n$ and $t$. For $\delta_T \gg 1$, a further approximation, $\mathcal{Z}_n \approx 1+e^{-\beta \mu_n(t_n^c)}$,
results in at most $e^{-\delta_T}$ relative error for each $P_n$, leading to a simple expression for the low-dispersion limit of the sudden decoupling mechanism,
\begin{align} \label{eq:gradtofit}
   P_n & = \bar{f}\bm{\left[}\mu_{n\!+\!1}^{}(t_{n+1}^c)\bm{\right]} \prod_{m=1}^{n} f\bm{\left[}\mu_m(t_{m}^c)\bm{\right]} \, .
\end{align}
The distribution \eqref{eq:gradtofit} is determined by a set of dimensionless numbers $\tilde{\mu}_{n}^{}\equiv \beta \mu_{n}^{}(t_{n}^c)$ and is
narrowly dispersed if $\ldots \gg \tilde{\mu}_{m+1} \gg \tilde{\mu}_{m} \gg \ldots$.

Now we consider the athermal limit. At sufficiently low temperatures the timescale for $f\bm{(}\mu_n(t) \bm{)}$ switching between loading ($\approx 1$) and unloading ($\approx 0$) may become much shorter than the time-scale of reducing $\Gamma_n(t)$. Assessing this gradual decoupling limit amounts to replacing
the Fermi functions in Eq.\eqref{eq:generalKinetic} by sharp steps,
$f(\mu_n) \to \Theta(t-t_{n}^{b})$. 
Starting with a sharp initial equilibrium free of thermal fluctuations, $P_n(t_0) =\delta_{n,N}$ with  $t_b^{N+1}< t_0< t_b^{N}$, the system of equations Eq.\eqref{eq:generalKinetic} is reduced to a set of decay cascade equations:
\begin{equation} \label{eq:generaldecay}
    \dot{P}_n(t) = \begin{cases}
      0, &  t< t_{n+1}^{b},\\
      \Gamma_{n+1} P_{n+1}(t), & t_{n+1}^{b} < t < t_{n}^{b},\\
      -\Gamma_{n} P_{n}(t)+ \Gamma_{n+1} P_{n+1}(t), &     t_{n}^{b} < t,
      \end{cases}
\end{equation}
with $P_0(t)$ set by normalization.
These equations generalize the decay cascade model~\cite{kaestner2010a} to distinct $t_n^{b}$'s.
One can show that a universal solution to Eqs.~\eqref{eq:generaldecay} independent of the specific shape of $\Gamma_n(t)$'s time-dependence~\cite{kaestner2010a},
\begin{align} \label{eq:decaycascade}
      P_{n} & = e^{ -X_{n}} \prod_{j=n+1}^{N} \left( 1 - e^{-X_{j}} \right),
\end{align}
remains valid in the limit of $\ldots \gg X_{j+1}\!\gg\!X_{j} \gg \ldots$ with appropriately generalized integrated decay rates, $X_n \equiv \int_{t_n^b}^{t_f} \Gamma_n(t) dt \stackrel{T\to0}{=}\int_{t_0}^{t_f} W^{-}_n(t) dt$.

\begin{figure}[t!] \centering
\centering \includegraphics{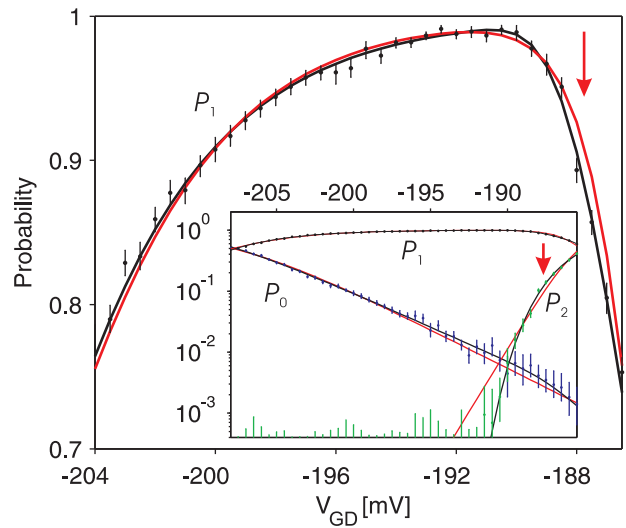}
\caption{(color online). Measured $P_1$ (points) as function of $V_\mathrm{GD}$, compared to theoretical fits to thermal and athermal limit (red and black solid lines, respectively). The error bars indicate 95~\% confidence intervals. The blue, black and green data points correspond to $P_0$, $P_1$ and $P_2$, respectively, shown on a logarithmic scale in the inset. 
}\label{fig:main}\end{figure}

Next we compare the two theoretical limits \eqref{eq:gradtofit} and \eqref{eq:decaycascade} with our measurement of $P_n$, as shown in Fig.~\ref{fig:main}. In the main diagram the experimental data for $P_1$ are plotted as symbols as function of $V_\mathrm{GD}$.
As indicated in Fig.~\ref{fig:principle}(b), $V_\mathrm{GD}$ controls $\tilde{\mu}_n$ and $X_n$. We assume $\tilde{\mu}_n $ to be linearly dependent on $V_\mathrm{GD}$ while typically decay rates depend exponentially on the gate voltage. Thus we use

\begin{subequations}\label{eq:VG}
\begin{align}
  \tilde{\mu}_n & = -\alpha_{\mu, n} V_\mathrm{GD} +\Delta_{\mu, n} \, \\
  \ln X_n & = -\alpha_{X, n} V_\mathrm{GD} +\Delta_{X, n},
\end{align}
\end{subequations}
incorporating $V_\mathrm{GD}$ dependence in Eqs.~\eqref{eq:gradtofit} and~\eqref{eq:decaycascade} using $\alpha_{\mu, n}$ and $\Delta_{\mu, n}$ as fit parameters for the thermal model and $\alpha_{X, n}$ and $\Delta_{X, n}$ for the athermal model. The red and black solid lines in Fig.~\ref{fig:main} show the result of fitting the thermal [Eq.\eqref{eq:gradtofit} and (\ref{eq:VG}a)] and athermal distribution [Eq.\eqref{eq:decaycascade} and (\ref{eq:VG}b)] to the experimental data around $P_1 \approx 1$. The blue, black and green data points correspond to $P_0$, $P_1$ and $P_2$ respectively, shown on a logarithmic scale in the inset. Only $0\!\leftrightarrow\!1$ and $1\!\leftrightarrow\!2$ transitions have been considered since $P_n$ with $n > 2$ do not contribute for $V_\mathrm{GD} < -186\,$mV (see Fig.~\ref{fig:probs}). The error bars on the data indicate 95~\% confidence intervals for estimating probability from the binomial statistics of direct counting~\cite{numRecipes} (see Appendix).
The thermal limit (red solid line) deviates beyond confidence interval for voltages $V_\mathrm{GD} > - 190\,$mV for $P_1$ on linear scale (see red arrow). Note also that the uncertainties for the model parameters turn out  larger for the thermal model (see Table~\ref{fitparams}). Even stronger deviations can be observed for $P_2$ on the logarithmic scale. Here the thermal model predicts a linear characteristic, while the apparent non-linearity is reproduced well by the athermal limit. In fact, the non-linear characteristic of $P_2$ is a direct signature of simultaneous variation in tunneling rates and energy during the capture process. Thus we conclude that our measured distribution is consistent with the generalized decay cascade model in the low-noise limit.

\begingroup
\squeezetable
\begin{table}
\caption{\label{fitparams} Best fit parameters.}
\begin{ruledtabular}
\begin{tabular}{|l|r|r|r|r|}
&\mbox{$\alpha_{\mu/X,1} /\text{mV}^{-1}$} & \mbox{$\alpha_{\mu/X,2} /\text{mV}^{-1}$} &\mbox{$\Delta_{\mu/X,1}$} &\mbox{$\Delta_{\mu/X,2}$}  \\ \hline
Thermal & $-0.293\pm 0.003$ & $-0.983\pm 0.023$ & $-60.8\pm 0.6$ & $-182.1\pm 4.2$\\ \hline
Athermal& $0.261\pm 0.003$ & $0.385\pm 0.009$ & $-54.5\pm 0.6$ & $-71.5\pm 1.6$\\ 

\end{tabular}
\end{ruledtabular}
\end{table}
\endgroup

The generalized decay cascade limit implies that lowering the lead temperature will not increase the initialization precision, which has indeed been found in surface acoustic wave driven devices~\cite{Janssen2001}. This feature alone would, however, not suffice to exclude the thermal distribution as in the past this saturation has been related to rf-heating induced by the modulation~\cite{Janssen2001}. Our results indicate a path for further improvements for dynamic QD initialization: it may be achieved by increasing the separation of decay steps ($X_n/X_{n-1}$) either by a large decay rate ratio ($\Gamma_n(t)/\Gamma_{n-1}(t)$) \cite{kaestner2010a} or sufficiently large energy separation $\Delta \mu$,
in which case $\Gamma_n(t^b_n)/\Gamma_{n-1}(t^b_{n-1})$ can be large even if $\Gamma_n(t) \approx \Gamma_{n-1}(t)$ due to the difference in $t^b_{n-1}$ and $t^b_{n}$. Alternatively, devices with reduced coupling between barrier and plunger may benefit from the \emph{thermal} limit which at sufficiently low temperature will further enhance the initialization precision. This regime may be reached by adding compensation pulses to $V_\mathrm{GD}$ during the transition (i) $\rightarrow$ (ii) in Fig.~\ref{fig:principle}. This example illustrates the importance of identifying the capturing mechanism for an efficient optimization of the initialization process of dynamic QDs as
building blocks for nanoelectronic circuits.

This work has been supported by the DFG.
V.K., J.T. and P.N. have been supported by
ESF project No.~2009/0216/1DP/1.1.1.2.0/09/APIA/VIAA/044.

\appendix
\renewcommand{\thepage}{S-\arabic{page}}
\renewcommand{\thefigure}{S\arabic{figure}}

\section{Appendix: supplemental material on data evaluation}

Here we detail the extraction of probabilities for the number of electrons $n$ which are captured on the quantum dot, subsequently transferred onto the node and observed by the detectors labeled Det1 and Det2 in Fig. 1 in the main text.

\begin{figure}[htb]
\centering
\includegraphics[width=0.65\linewidth]{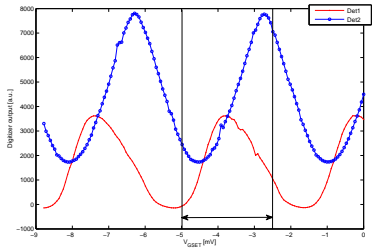}
\caption{Typical SET responses under variation of the SET gates. The red trace corresponds to Det1, the blue one to Det2. The vertical black lines indicate the range of gate voltage settings used for working point variation of the SETs.}
\label{SETvsGate}
\end{figure}

In contrast to other detection methods, like e.g. quantum point contacts, the superconducting SETs used here to monitor the node's charge state respond periodically to external potentials thus showing a specific response (amplitude as well as direction) depending on the SET's working point. This is influenced by all potentials in close vicinity to the SET island via capacitive coupling. In addition to the gates controlling the semiconducting part of the chip by applying voltages $V_\text{GS}$, $V_\text{GD}$ and $V_\text{barrier}$ we have evaporated gates close to each of the SET islands (referred to in the following as SET gates with voltages $V_\text{GSET}$) which allow for the control of the SET working points independently from any other potential in the circuit. The gate voltage $V_\text{GSET}$ is always the same for both SET gates. A typical current response for both detectors under variation of $V_\text{GSET}$ is shown in Fig. \ref{SETvsGate}. For each value of $V_\text{GD}$ (gate between the quantum dot and the node), we change the voltage $V_\text{GSET}$ (and thereby the SET working points) in 20 steps over half a SET period. The parameter range is indicated by the black vertical lines in Fig. \ref{SETvsGate} which corresponds to $V_\text{GSET}\in [-2.5,5]$ mV.

\begin{figure}[htb]
\centering
\includegraphics[width=0.6\linewidth]{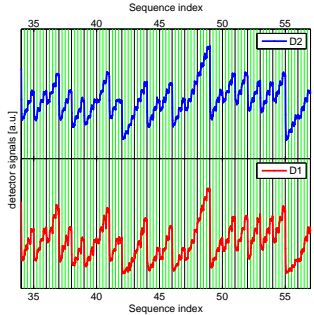}
\caption{Signal traces for Det1 (lower graph, red trace) and Det2 (upper graph, blue trace) both measuring the same signal. Vertical lines indicate pulse positions as explained in the text.}
\label{trace_raw}
\end{figure}

For each setting of the static gates $V_\text{GD}$ and $V_\text{GSET}$ we now take a timetrace of the detectors' current response to a specific pulse sequence (see Fig.~\ref{trace_raw}): We first apply a short pulse to the gate $V_\text{barrier}$ between the node and the drain lead (black vertical line) which randomly changes the number of electrons stored on the node close to an equilibrium number. Due to this pulse, we avoid the buildup of a strong electrostatic potential on the node which on the one hand would couple capacitively to the quantum dot (mesoscopic feedback) but would also potentially drive the SETs out of sensitivity. After this first pulse, three combined capture and transfer pulses onto gate $V_\text{GS}$ of the dynamic quantum dot are applied which are illustrated by green vertical lines in Fig. \ref{trace_raw}. All pulses are delayed by $t_\text{delay}=40$ ms each. Every timetrace consists of 279 sequence periods, i.e. 279 reset pulses and 837 combined capture and transfer pulses. 

\begin{figure}[htb]
\centering
\includegraphics[width=.8\linewidth]{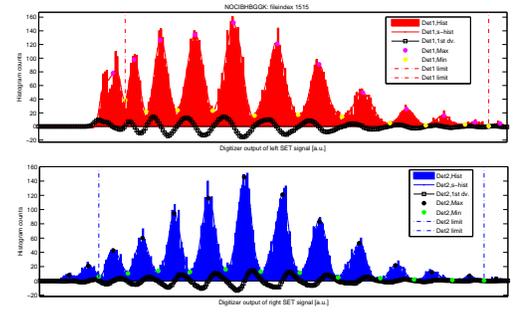}
\caption{Histograms for both traces together with the first smoothed derivative and the maxima/minima found indicated by the dots. The vertical dashed lines show the limits of sensitivity set by the first and last minimum found.}
\label{hist}
\end{figure}

\begin{figure}[htb]
\centering
\includegraphics[width=.8\linewidth]{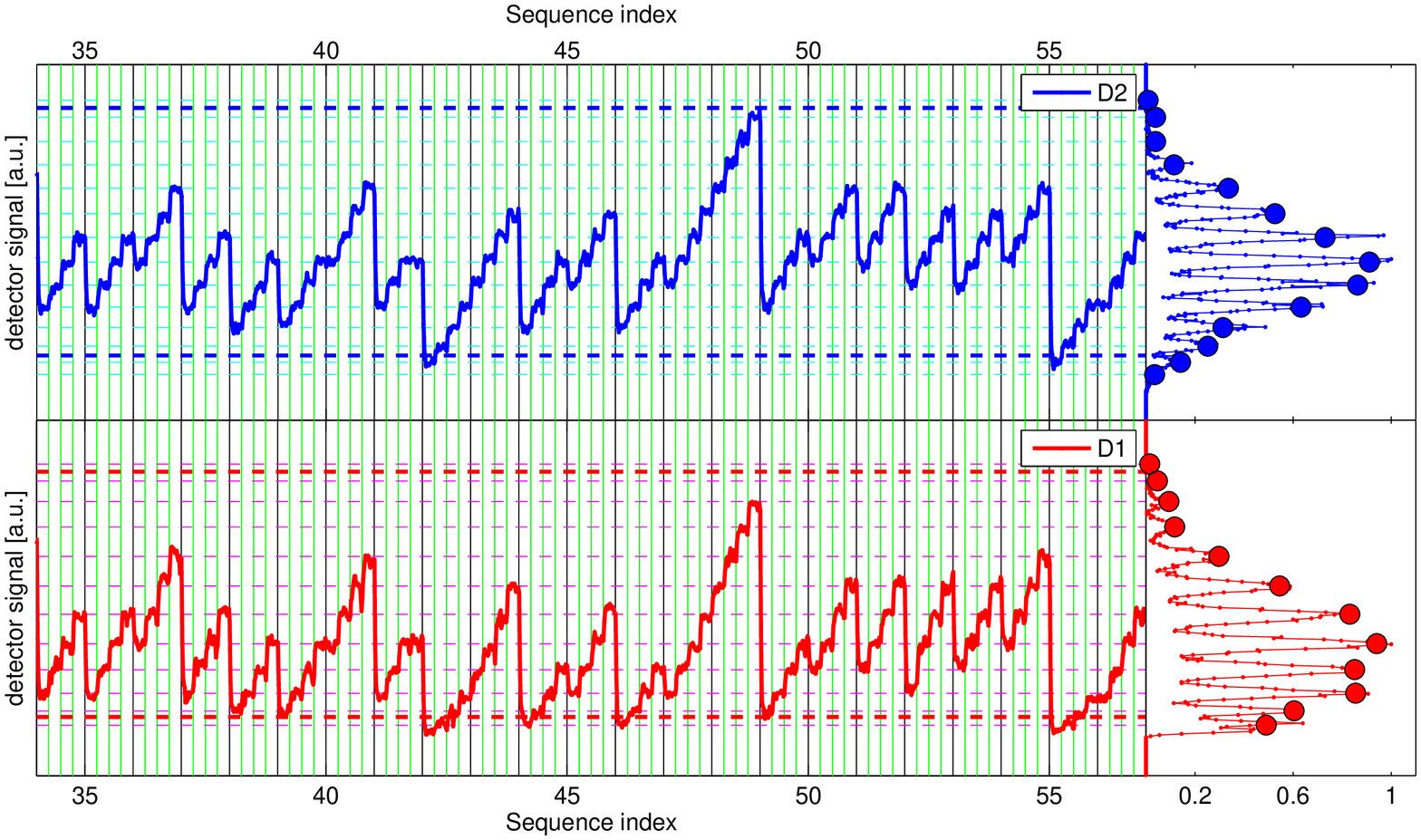}
\caption{Signal traces together with the histogram and the levels found (dashed horizontal lines in the trace) and the range of SETs' sensitivity (bold dashed lines).}
\label{trace}
\end{figure}

Next we describe an example of evaluating the trace shown in Fig.~\ref{trace_raw}:
We first make a histogram (see Fig.~\ref{hist} and \ref{trace}) of the first 175 sequence periods of the two time traces, then apply a second-order Savitzky-Golay filter in order to smooth the data and calculate the derivative (black lines), which yields the maxima and minima of the histograms of the time traces. The maxima found correspond to the occupation levels of the node. The data range of the SETs indicating the range of sensitivity is also derived from this calculation and is limited by the first and last minimum found by the algorithm. These either mark the vanishing separation of levels close to the SET extrema (lower limit for Det1) or possibly insufficient occupation of the following levels for reliable detection (upper limit for Det1, both limits for Det2).

The resulting levels are then compared to each mean value of the corresponding detector trace between two pulses. Based on the assumption that each possible charge state of the node is found by the level detection (which is supported by the stochastic reset of the node's charge number in each sequence) we can directly extract the number of charge carriers between two pulses by counting the number of levels lying in-between the starting and the final level. Whenever the starting or the final state is beyond these limits set by the algorithm (i.e., the SET is not sensitive) the corresponding event is not considered in the statistics. 

This evaluation is performed for every parameter set $(V_\text{GD},V_\text{GSET})$. The results are combined for such traces where $V_\text{GD}$ is constant, both of the detectors are sensitive for at least 30\% of the time and more than 150 equal events for both detectors have been observed. The last argument is based on the fact that both detectors are coupled to the same island and therefore the two signals are not independent but correlated in terms of electron number difference between two pulses. By applying this restriction we can strongly reduce the number of erroneous counts caused by noise on one of the detectors or by failures in the algorithm (mainly errors in the level detection). All events or traces not matching these criteria, e.g. pulses at insensitive working points for at least one SET or inequal results for both detectors, are neglected. Therefore, the total number of events $N(V_\text{GD})$ evaluated for one setting of $V_\text{GD}$ varies from 1853 to 8997 events for $V_\text{GD}\in [-230,-175]$ mV (average number of events $\approx 6100$).

Finally, this procedure results in a sequence of electron numbers captured and transferred by the pulses applied to the dynamic QD. 
For each $V_\text{GD}$ we count the number of outcomes $M_n(V_{\text{GD}})$ that have resulted in exactly $n=0\ldots 4$ electrons captured and transferred in one cycle. $M_n(V_{\text{GD}})$ is expected to obey the binomial distribution with the (unknown) success probability 
$P_n(V_{\text{GD}})$  and the total number of trials $N(V_\text{GD})$.
The probability $P_n$ is then estimated by definition as success frequency: $\bar{P}_n = M_n/N$.
We have used the principle of maximal likelihood and the 
quantile function of the
binomial distribution to estimate the (symmetric) $p=95 \%$ confidence intervals for  $P_n$,
as depicted by error bars in Fig.~3 and 4.
For the special outcomes of $M=0$ and  $M=N$ the confidence interval is estimated by requesting the expected probability of this extremal outcome 
to be no less than $p$.

For the best-fit estimation of parameters of the models expressed by Eq. (6) and (9a) or Eqs. (8) and (9b) we
have used a standard  least-squares routine, matching the Gaussian model it assumes to  $p=68 \%$ confidence intervals obtained from the exact binomial statistics
as described above. The confidence intervals for the model parameters are then computed using Monte Carlo simulation of 
$500$ synthetic data sets for each of the two models separately. The synthetic data are drawn from binomial distributions
with success probabilities  $P_n(V_\text{GD})$  taken from the respective best-fit model and the number of 
trials  $N(V_{\text{GD}})$ equal to that in the actual experiment for each $V_{\text{GD}}$.

\end{document}